\documentclass[%
 preprint,
 amsmath,amssymb,
 aps, physrev,
]{revtex4-2}

\usepackage{graphicx}
\usepackage{dcolumn}
\usepackage{bm}

\begin{document}

\preprint{APS/123-QED}

\title{\textbf{Magnetoelastic coupling at the field-induced transition in EuAl$_{12}$O$_{19}$} 
}%

\author{T. Haidamak}
\email{Contact author: tetiana.haidamak@matfyz.cuni.cz}
 \altaffiliation{}
\author{G. Bastien}%
\author{P. Proschek}%
\author{A. Eliáš}%
\author{R.H. Colman}%

\affiliation{%
 \\
  Department of Condensed Matter Physics, Faculty of Mathematics and Physics, Charles University, Ke Karlovu 3, 121 16 Prague 2, Czech Republic}%
\author{D. Gorbunov}
\author{S. Zherlitsyn}
\affiliation{
 Hochfeld-Magnetlabor Dresden (HLD-EMFL) Helmholtz-Zentrum Dresden-Rossendorf (HZDR), 01328 Dresden, Germany}%

\author{A.A. Zvyagin}
\affiliation{
 B. Verkin Institute for Low Temperature Physics and Engineering of the National Academy of Sciences of Ukraine, Nauky Ave., 47, Kharkiv, 61103, Ukraine}%
\affiliation{ Max-Planck Institut fur Physik komplexer Systeme, Nothnitzer Str., 38, D-01187, Dresden, Germany}%
\author{G.A. Zvyagina}
\affiliation{%
B. Verkin Institute for Low Temperature Physics and Engineering of the National Academy of Sciences of Ukraine, Nauky Ave., 47, Kharkiv, 61103, Ukraine\\}%
\author{J. Prokleška }
\author{V. Sechovský}
\author{M. Vališka}
\affiliation{
 Department of Condensed Matter Physics, Faculty of Mathematics and Physics, Charles University, Ke Karlovu 3, 121 16 Prague 2, Czech Republic}%

\date{\today}

\begin{abstract}

Magnetoelastic coupling plays a crucial role in magnetic-field-induced transitions in anisotropic ferromagnets. Ultrasonic methods are suitable for experimental investigations of these phenomena. We investigate elastic constants in EuAl$_{12}$O$_{19}$, a quasi-two-dimensional anisotropic ferromagnet, by measuring sound velocity in magnetic fields perpendicular to spontaneous magnetization. The shear modulus $C_{44}$ exhibits dramatic softening at the field-induced transition from the ferromagnetic to a paramagnetic phase with magnetic moments forced to polarize along the applied transverse field. The softening is attributed to strong magnetic fluctuations near a second-order phase transition. Theoretical calculations based on magnetization data qualitatively reproduced the observed behavior within a strain-exchange mechanism. These results demonstrate that magnetoelastic coupling in EuAl$_{12}$O$_{19}$ arises primarily from exchange striction and provide a framework for modeling similar transitions in other anisotropic ferromagnets.

\end{abstract}
\maketitle
\section{Introduction}

Knowledge of the interaction between magnetic moments and the crystal structure is essential for understanding magnetism in solids. In particular, magnetoelastic coupling (MEC) stabilizes the magnetic order in a quasi-two-dimensional ferromagnet  \cite{Mitsay}  and influences the domain structure \cite{McCray}. 

Here, we consider the simple problem of an anisotropic ferromagnet under a magnetic field applied perpendicular to the direction of spontaneous magnetization. In this case, the two phases in the competition are (a) the ferromagnetic phase (FM) with the spontaneous magnetization component and a perpendicular magnetization component induced by the external magnetic field, and (b) the paramagnetic phase (PM) with magnetic moments forced by the magnetic field to align along its direction. The transition is usually of the second-order type (SOMPT). At sufficiently low temperatures, well below $T_{C}$, it can be of first-order type  (FOMPT) and transforms the ferromagnet into a polarized paramagnet (PPM) with magnetic moments aligned with the field \cite{Melville, Asti, Barbara_1978, ASTI198029}. The position of the tricritical point in the magnetic phase diagram where SOMPT and FOMPT meet depends on the strength of the FM exchange interaction and magnetocrystalline anisotropy.   

The FOMPT from FM to perpendicular PPM phase was intensively studied in URhGe in connection with a transverse-field-induced reentrant superconductivity \cite{Levy,Brando,Knafo2012}, and in U$_{3}$Cu$_{4}$Ge$_{4}$ \cite{Gorbunov}. 

In the second case, the transition reaches 0 K at a finite magnetic field, and thus, it becomes a first-order quantum phase transition \cite{Gorbunov,Brando, Mineev}. The behavoir of an anisotropic ferromagnet under a transverse magnetic field was also investigated in insulating or semiconducting ferromagnets such as the ferromagnetic spin chain CoNb$_{2}$O$_{6}$  \cite{Coldea}, the tetragonal ferromagnet LiHoF$_{4}$ \cite{Bitko}, the honeycomb ferromagnets Cr$_{2}$X$_{2}$Te$_{6}$ (X = Si, Ge) \cite{Selter,Spachmann,Zhang} and the triangular lattice ferromagnets NaBaCoV$_{2}$O$_{8}$ \cite{Nakayama} and EuAl$_{12}$O$_{19}$ \cite{BastienPRB}. In these cases, the SOMPT has only been reported based on specific heat \cite{BastienPRB}, neutron diffraction  \cite{Coldea}, or magnetostriction data \cite{Spachmann,Zhang}. In  CoNb$_{2}$O$_{6}$ \cite{Coldea} and LiHoF$_{4}$ \cite{Bitko} the transition shifts to lower temperatures with increasing field and reaches 0 K at a quantum critical point. This behavior can be applied in magnetic refrigeration \cite{Wolf}, as experimentally evidenced in the case of LiHoF$_{4}$ \cite{Liu}. 

The nature of the transverse-field-induced magnetic phase transition is expected to be influenced by the related lattice properties. Such coupling was previously revealed in several anisotropic ferromagnets under a transverse magnetic field by thermal-expansion measurements \cite{Nikitin, Spachmann}, magnetostriction measurements \cite{Gorbunov,Spachmann} or by the measurement of the hydrostatic or uniaxial pressure dependence of the field-induced magnetic transition \cite{Miyake,Braithwaite,Olmos}. The influence of magnetoelastic coupling can be investigated more thoroughly by measuring variations in sound velocity in magnetic fields and determining the behavior of corresponding elastic constants.

Measurement of sound velocity is a powerful technique for understanding magnetoelastic effects. Specifically, it has brought new insights into various magnetoelastic effects in magnetic materials including field-induced transitions in an uniaxially anisotropic ferromagnet U$_{3}$Cu$_{4}$Ge$_{4}$ \cite{Gorbunov}, an antiferromagnet UIrSi$_{3}$ \cite{Gaid2022}, a ferrimagnet Dy$_{2}$Fe$_{14}$Si$_{3}$ \cite{Andreev}, itinerant metamagnets UTe$_{2}$ \cite{Valiska} and UCoAl \cite{Yoshizawa}.

EuAl$_{12}$O$_{19}$ is ideally suitable for a sound velocity study of the transverse-field-induced phase transition, considering the possibility of growing single crystals of several mm$^{3}$ by the floating zone method \cite{BastienPRB}. It is also a much simpler case than U$_{3}$Cu$_{4}$Ge$_{4}$ since the Eu$^{2+}$-ion moments in EuAl$_{12}$O$_{19}$ are localized and spin-only ($S$ = 7/2 and $L$ = 0) \cite{BastienPRB} similar to Gd$^{3+}$ ions. EuAl$_{12}$O$_{19}$ is a quasi-two-dimensional ferromagnet with  $T_{C}$ = 1.3 K \cite{BastienPRB}. The Eu$^{2+}$ ions form planar triangular lattices within the hexagonal magnetoplumbite crystal structure (space group $P63/mmc$) \cite{BastienPRB,BastienAdvMat}. The single-ion anisotropy of the Eu$^{2+}$ ions, leads to the spontaneous magnetization oriented along the $c$-axis. The magnetic field $\mu_{0}H$ = 0.3 T applied perpendicular to the easy magnetization axis lowers $T_{C}$ to 1 K \cite{BastienPRB}.

In this paper, we investigate the elastic response of EuAl$_{12}$O$_{19}$ at the phase transition induced by applying a magnetic field transverse to the easy magnetization axis by measuring isotherms of relative sound velocity change ($\Delta v$/$v$) in varying magnetic fields. To better monitor the phase transition, we also measured the corresponding magnetic-field dependencies of magnetization and specific heat. The experimental results are compared with the theoretical calculations based on a strain-exchange mechanism using magnetization data. The obtained results confirm the second-order nature of the magnetic transition.

\section{Experimental}
High-quality EuAl$_{12}$O$_{19}$ single crystals prepared by the optical floating zone method by Bastien et al. \cite{BastienPRB} were used for our experiments. We measured magnetization and specific heat 
on the same samples that Bastien et al. measured. Specific-heat measurements were performed using the relaxation method employing a two-tau model. Owing to the low thermal conductivity of the crystals and the consequently long relaxation times, a long-pulse method was also employed. The sample, used for the ultrasound study with dimensions of $1.45\times  2.61\times0.92$ mm, for the $a$-, $a^*$- and $c$-axes, respectively, has been cut from the same batch. The samples were oriented by X-ray diffraction via the Laue method.

Hook’s law $\sigma_{\mathrm{ij}}$ = $C_{\mathrm{ijkl}}$$e_{\mathrm{kl}}$, describes the relations between stress $\sigma_{\mathrm{ij}}$ and strain $e_{\mathrm{kl}}$ tensors and elastic constant tensor $C_{\mathrm{ijkl}}$. The hexagonal EuAl$_{12}$O$_{19}$ has five independent elastic constants ($C_{11}$, $C_{33}$, $C_{44}$, $C_{12}$, $C_{14}$) \cite{luthii}. Only the diagonal elements of the elastic constants matrix can be measured directly, so we have measured the temperature- and magnetic-field-dependencies of $C_{11}$ and $C_{44}$.  Each of the constants corresponds to a specific direction of elastic wave propagation. There are two types of elastic constants: longitudinal, where $q$-wave and $u$-displacement vectors are collinear; and transverse, where $q$-wave and $u$-displacement vectors are perpendicular. In the case of europium hexaaluminate, $C_{11}$ ($q$ll$a$, $u$ll$a$) and $C_{33}$ ($q$ll$c$, $u$ll$c$) are examples of longitudinal modes, and the shear mode $C_{44}$ ($q$ll$a^*$, $u$ll$c$) is considered transverse.

Usually, longitudinal elastic constants are less sensitive to magnetoelastic interactions compared to transverse ones \cite{luthii}. But for example in two-dimensional dimer system SrCr${}_{2}($BO$_{3}){}_{2}$ longitudinal mode $C_{11}$ shows strong temperature and field dependence. Such behavior was due to strong inter-dimer spin-strain coupling. At low temperatures strong softening in both sound velocity and magnetization was observed \cite{Zherlitsyn62}. 

In the majority of cases, the transverse modes, such as $C_{44}$ and $C_{66}$, are most fruitful in terms of anomalies. For example, in the case of magnesium chromite spinel MgCr${}_{2}$O${}_{4}$  the tetragonal shear modulus shows significant softening in temperature, as well as the trigonal shear modulus $C_{44}$, which exhibited non-monotonic temperature dependence \cite{Watanabe}. In  geometrically frustrated spinel ZnFe${}_{2}$O${}_{4}$ the frustration effects from the viewpoint of spin-lattice coupling were studied \cite{Takita_2015}. The so-called soft-mode $C_{66}$ usually exhibits strong softening up to a few percent. However, significant softening, up to 25 percent in high external magnetic fields up to 50 T and up to 2.5 percent in temperature scans, was observed in MgCr${}_{2}$O${}_{4}$ \cite{luthii,Watanabe}. Such sensitivity of shear modes could be observed in both temperature- and field-dependent measurements, reflecting coupling among elastic, magnetic, and electronic subsystems. To clarify interactions of a system, it is important to investigate as many independent constants as possible, but at least one of the longitudinal and one of the transverse modes.
The higher the symmetry of the crystal system, the fewer independent constants it has \cite{luthii}. In our case, the hexagonal EuAl$_{12}$O$_{19}$, implies only one independent transverse mode, the shear mode $C_{44}$ \cite{luthii}. $C_{44}$ mode matches relation between the $c$-axis and $aa^*$-plane, and the longitudinal $C_{11}$ mode reflects interaction along the $a$-axis. We observed anomalies in the field dependencies of $C_{44}$ mode at various temperatures. The $C_{11}$ mode appeared entirely intact by the phase transition. No anomalies were detected in the signal noise. Further, solely acoustic data related to the $C_{44}$ mode will be presented and discussed.

The phase-sensitive detection technique was used to study the elastic properties of EuAl$_{12}$O$_{19}$ \cite{luthii,Kohama}. The single crystal was polished with the faces perpendicular to the principal crystallographic axes. Ultrasound waves were generated by LiNbO$_{3}$ transducers glued directly onto the sample with Thiokol LP-032 glue. We used two types of transducers, depending on the measured mode (longitudinal or transverse). 

\begin{figure}
\begin{center}
\includegraphics[scale=1]{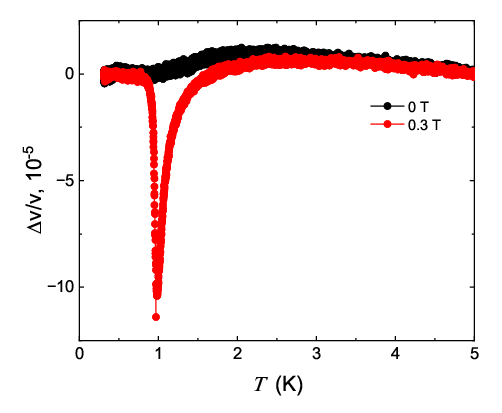}
\end{center}
\caption{Temperature dependence of $\Delta v$/$v$ of $C_{44}$ mode in EuAl$_{12}$O$_{19}$ in a magnetic field of 0  and 0.3 T, respectively, applied along the $a^*$-axis.}
\label{Fig.1}
\end{figure}

Bastien et al. \cite{BastienPRB} studied isothermal magnetization curves and isofield temperature dependencies of specific heat in magnetic fields applied along and perpendicular to the easy magnetization axis. Data collected in transverse fields serve as a useful reference for our study. From the latter experiment, they concluded that the transverse fields lower the Curie temperature $T_{C}$. $T_{C}$ = 1 K was determined at $\mu_{0}H$ = 0.3 T.  In Fig.~\ref{Fig.1}, we show the temperature dependence of $\Delta v$/v of the $C_{44}$ mode in zero magnetic field and in a transverse field of 0.3 T. In a zero field, a shallow valley is seen around $T_{C}$. A sharp inverted $\lambda$ -shape dip with a minimum at 1 K is observed in the field of 0.3 T. 

The isothermal dependencies of magnetic fields up to 10 T were measured at temperatures from the lowest achievable temperature, 0.3 K, to 1.5 K, which is just above $T_{C}$. Fig.2c shows plots in fields up to 1 T, sufficient to illustrate variations of the anomaly associated with the transverse-field-induced phase transition field dependencies at temperatures down to 0.3 K. At temperatures below 1.2 K, the anomaly has an inverted-$\lambda$-like shape, with the minimum pointing to the critical field of the transition. It increases and moves to higher fields as the temperature decreases. At 1.2 K, it is dramatically reduced and vanishes above $T_{C}$. The magnetization curves likely exhibit a break (a sudden drop in the first derivative) at $H_{c}$ since it coincides with the minimum of $\Delta v$/$v$ observed at the same temperature. The corresponding field dependencies of specific heat at $T \geq 0.8 K$) drop near the envisaged $H_{c}$. The $Cp/T$ vs. $H$  data for $T \leq 0.5 K$ are likely hampered by extremely slow relaxations due to the very low heat conduction of EuAl$_{12}$O$_{19}$ and are not considered for the analysis.

\begin{figure}
\begin{center}
\includegraphics[scale=.75]{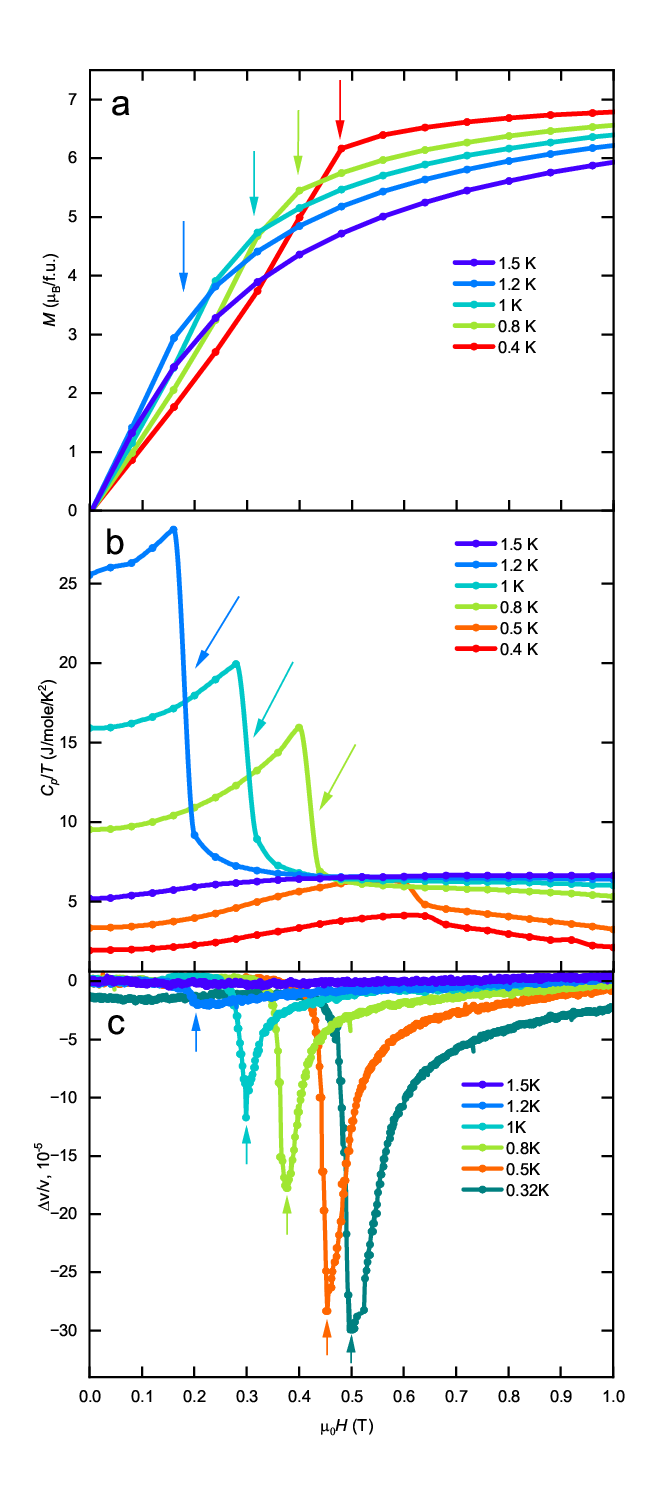}
\end{center}
\caption{a) magnetization, b) heat capacity, and c) $\Delta v$/$v$ of $C_{44}$ mode dependencies on the magnetic field applied along the $a^*$-axis. The arrows point to $H_{c}$.}
\label{Fig.2}
\end{figure} 

The $H-T$ phase diagram constructed from $H_{c}$ values derived from our isothermal magnetic-field dependencies of magnetization, specific-heat, and $\Delta v$/$v$ of $C_{44}$ shows reasonable agreement with the values derived from isofield temperature dependencies of specific-heat measured by Bastien et al.\cite{BastienPRB}. The $H_{c}$ values derived from the corresponding magnetization curves (also shown in the figure) are significantly lower. This discrepancy, which likely originates from the limited sensitivity of the measurement technique used in Ref. \cite{BastienPRB}, will be discussed later. Nevertheless, the overall agreement between the various methods clearly defines the phase boundary between the FM and PM phases, allowing us to discuss the characteristic signatures of the employed methods and to determine the order of the corresponding transition.

\begin{figure}
\begin{center}
\includegraphics[scale=1]{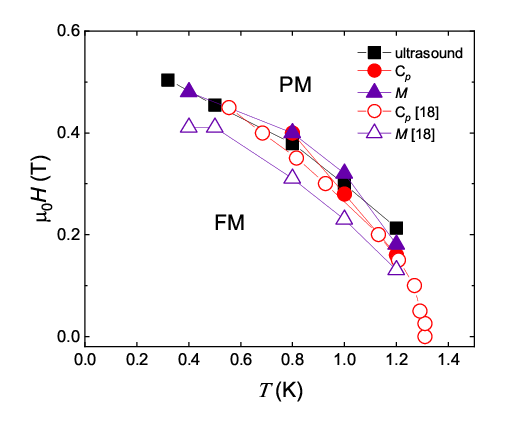}
\end{center}
\caption{$H-T$ phase diagram of EuAl$_{12}$O$_{19}$ in the magnetic field applied along the $a^*$-axis. The transition data points derived from our ultrasound - full squares, specific heat - full circles, and magnetization data - full triangles, and complementary specific heat - open circles and magnetization data - open triangles published by Bastien et al.\cite{BastienPRB}.}
\label{Fig.3}
\end{figure}  

\section{Theory}

For the strain-exchange mechanism, the sound-velocity $\Delta v$ renormalization manifests itself as follows \cite{TM}. Sound waves change the positions of magnetic and nonmagnetic ions, which are involved in the direct exchange or indirect (superexchange) interaction between magnetic ions. In turn, spin-spin interactions renormalize the sound parameters, especially the sound velocity. Let us define the functions \cite{TM}:
\begin{eqnarray}
&&g({\bf q}) = \sum_j e^{i{\bf q}{\bf R}_{ji}}\left( e^{i{\bf k}{\bf R}_{ji}} -1\right) {\bf u}_{\bf k} \frac {\partial J_{ij}^{\alpha \beta}}{\partial {\bf R}_i} \ , \nonumber \\ 
&&h({\bf q}) =\sum_j e^{-i{\bf q}{\bf R}_{ji}}\left[ 1-\cos ({\bf k}{\bf R}_{ji})\right] \times \nonumber \\ 
&&\times ({\bf u}_{\bf k}\cdot {\bf u}_{-{\bf k}})\frac {\partial^2 J_{ij}^{\beta,\beta '}}{\partial {\bf R}_i \partial {\bf R}_j} \ . 
\label{gh}
\end{eqnarray}
Here we use the following notations: ${\bf R}_{ji} = {\bf R}_j -{\bf R}_i$, with ${\bf R}_j$ are the position vector of the $j$-th site of the magnetic ion, and $J_{ij}^{\alpha \beta}$ ($\alpha \beta = x,y,z$) are the components of the exchange interaction between magnetic ions on the $i$-th and $j$-th site (we assume the anisotropic exchange). The parameters ${\bf k}$ and ${\bf u}_{\bf k}$ denote the wave vector and the polarization of the sound wave, respectively. The functions $g({\bf q})$ and $h({\bf q})$ are magneto-elastic coupling coefficients. 

The analysis \cite{Gen} shows that the renormalization of the sound velocity is proportional to the four-spin correlation function. The four-spin correlation functions can be approximated as combinations of two-spin correlation functions and average values of spin projections. The latter are connected with the magnetization per magnetic ion $M_i^{\alpha}$ ($\alpha = x,y,z$) via the expression $\langle S_i^{\alpha} \rangle = M_i^{\alpha}/\mu^{\alpha}$, where $\mu^{\alpha} =g_{\alpha} \mu_B$, with $\mu_B$ being the Bohr magneton and $g_{\alpha}$ being the effective $g$-factor of the magnetic ion for the direction $\alpha$. On the other hand, the average value of the pair spin-spin correlation function can be connected with the components of the spin susceptibility per magnetic ion $\chi^{\alpha \beta}({\bf q})$ ($\alpha, \beta = x,y,z$, ${\bf q} $ is the wave vector)
\begin{equation}
\frac{[\langle S_i^{\alpha}S_j^{\beta} \rangle -\langle S_i^{\alpha}\rangle \langle S_j^{\beta}\rangle ]}{k_BT}  = \frac {\chi^{\alpha \beta}({\bf q})}{\mu^\alpha \mu^{\beta} } ,
\end{equation} 
where $T$ is the temperature and $k_B$ is the Boltzmann constant.  

\begin{figure}
\begin{center}
\includegraphics[scale=1]{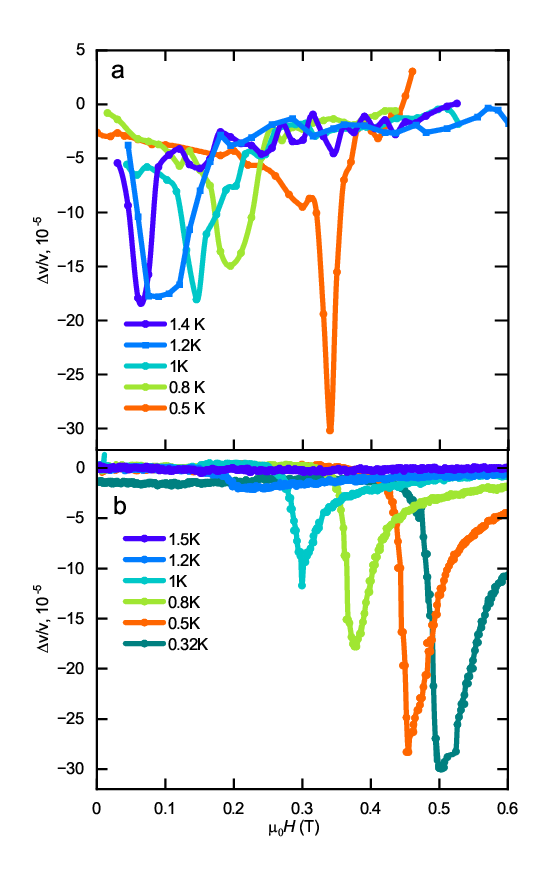}
\end{center}
\caption{$\Delta v$/$v$ related to $C_{44}$ mode in EuAl$_{12}$O$_{19}$  at selected temperatures as function of the external magnetic field $H$ directed along the $a^*$-axis a) calculated within the strain-exchange model using the magnetization data from Ref. \cite{BastienPRB}, b) determined by our ultrasound experiment.}
\label{Fig.4.png}
\end{figure}

In many works, where the renormalization of the sound characteristics of magnetic systems of various nature were studied in experiments and theoretically, e.g., for antiferromagnets with rare-earth magnetic ions \cite{And2017}, easy-axis antiferromagnets with transition metal and rare-earth magnetic ions \cite{Zv2008,Zv2010}, easy-plane antiferromagnets \cite{Zv2011}, and for ferrimagnets \cite{Andreev}, for 
U-based antiferromagnets \cite{Zv2012,UN,Gaid2022}, for gapped spin systems \cite{Nomura}, low-dimensional spin systems \cite{Sytcheva}, spin systems with frustration \cite{Zher2015}, including classical and quantum spin ices \cite{Erf2011,Bh2016}, and for systems with spin and orbital ordering \cite{Kol2017}, it was shown that it is possible to neglect the inhomogeneity of the spin susceptibility. As a result we obtain 
\begin{eqnarray}
&&\frac {\Delta v}{v} \approx  - \frac {1}{\rho V \omega^2\mu^4} 
\biggl[|g(0)|^2(2M^2\chi + \nonumber \\
&&k_BT\chi^2) + h(0)\mu^2(M^2 +k_BT\chi)\biggr], \
\label{dv}
\end{eqnarray}
where $V$ is the volume of the crystal, and $\omega$ is the frequency of the sound wave. The coefficients $g (0)$ and $h(0)$, see Eqs.~(\ref{gh}), can be used as fitting parameters. Using Eq.~(\ref{dv}) and the data of the low-temperature experiments for the magnetization of EuAl$_{12}$O$_{19}$ \cite{BastienPRB} we calculate the renormalization of the sound velocity. 

\section{Discussion}

In systems with spin-lattice interaction, the sound velocity is renormalized due to the magnetoelastic coupling \cite{luthii}. In crystals with enough high symmetry, components of the elastic modulus $C$ are coupled to the related velocities of sound $v$, via the formula $C = \rho v^2$, where $\rho$ is the density of the crystal. Within the single-ion mechanism, the renormalization of the elastic modulus is caused by the strains of non-magnetic ions (ligands), resulting from the sound wave \cite{luthii}. These strains change the crystalline electric field, which, together with the spin-orbit coupling, yields the single-ion magnetic anisotropy. Hence, the sound-caused strains of ligands result in a renormalization of the single-ion magnetic anisotropy. Vice versa, changes in the single-ion magnetic anisotropy of magnetic ions renormalize the elastic modulus, and hence, the sound velocities. Fig.~\ref{Fig.1} shows the features related to the phase transition in the behavior of sound velocities. The single-ion anisotropy itself cannot cause phase transitions. Hence, we use the strain-exchange scenario for the description of the changes of sound velocities in EuAl$_{12}$O$_{19}$. Notice that we used numerical differentiation of the magnetization to obtain the spin susceptibility. In that process, we  neglected noise to focus on the main contributions of the magnetization data changes as a function of the applied magnetic field $H$. The upper panel of Fig.~\ref{Fig.4.png} manifests the results of our calculations for the renormalization of the sound velocity related to the component of the elastic modulus $C_{44}$ as a function of the magnetic field $H$ directed along the $a^*$-axis of the crystal. The relative shifts between the fields at which corresponding anomalies have been theoretically and experimentally determined can be most likely attributed to the fact that the theoretical calculations were based on magnetization data taken from \cite{BastienPRB}, which reported considerably lower $H_{c}$ values than the specific-heat data presented in the same paper (see Fig. 5 in \cite{BastienPRB}) and results of magnetization, specific-heat, and $\Delta v$/$v$ measurements (see Fig.~\ref{Fig.2}  and Fig.~\ref{Fig.3}). The magnetization data presented in Ref. \cite{BastienPRB} were hampered by the low sensitivity of Hall probes, which were the only technique available at the time of those experiments. In the present work, magnetization measurements were performed using a SQUID magnetometer MPMS 3 (Quantum Design) equipped with a $^3$He option, providing several orders of magnitude higher sensitivity.

After analyzing the results, we can point out that:

First, the strain-exchange model qualitatively reproduces the magnetic field features and temperature behavior of the renormalized sound velocities in EuAl$_{12}$O$_{19}$. Hence, we can conclude that the strain-exchange mechanism mainly contributes to the magneto-elastic coupling in this compound. 

Second, we can estimate the magneto-elastic coupling constants in that system. Our estimations imply that $|g(0)|^2/V\rho \omega^2 \sim 10^{-7}$ and $h(0)/V\rho \omega^2 \sim 10^{-8}$, i.e., the coupling between the magnetic and the elastic subsystems in EuAl$_{12}$O$_{19}$ is relatively weak. 

Third, the renormalization of the sound velocity reveals dramatic softening of the elastic modulus $C_{44}$ at the magnetic phase transition in the system. 

The principal question that remains to be discussed and answered concerns the type of magnetic phase transition from the FM to the PM state induced in EuAl$_{12}$O$_{19}$ by a magnetic field transverse to spontaneous magnetization. As known from previous studies of other ferromagnets, such a transition is usually SOMPT; at sufficiently low temperatures, well below $T_{C}$, it can be FOMPT, transforming the FM into a polarized paramagnet (PPM) with magnetic moments aligned with the field\cite{Melville, Asti, Barbara_1978, ASTI198029, Levy, Gorbunov}. 
SOMPT and FOMPT can sometimes be difficult to distinguish experimentally. FOMPT is characterized by latent heat, which causes hysteresis of the anomalies in the measured property at the critical point. In the case of EuAl$_{12}$O$_{19}$, no hysteresis has been detected by measurements of specific heat and magnetization, neither by Bastien et al. \cite{BastienPRB} nor by us. However, this negative result does not entirely rule out the presence of FOMPT. A small hysteresis can be hidden within inaccuracies of a sub-Kelvin experiment. SOMPT in magnetic systems is usually accompanied by spin fluctuations that slow down enormously near the critical point, while FOMPT is generally devoid of fluctuations. 

At this point, the sound velocity measurements can demonstrate their usefulness in our study. The sound velocity in magnetic systems exhibits specific anomalies near the SOMPT critical point. Magnetoelastic coupling mediates a significant influence of magnetic fluctuations on the crystal lattice, leading to lattice instabilities that are often manifested as softening of elastic modes. The singular behavior of sound velocity due to anomalies in elastic modes reflects extreme fluctuations of the magnetic ordering parameter near the critical temperature. These fluctuations give rise to a characteristic deep minimum in the sound velocity at the critical point.

On the other hand, the $\Delta v$/$v(H)$ dependence exhibits just a single step at the FOMPT critical point. The size of the step corresponds to the difference in the characteristic $\Delta v$/$v(H)$  values for the two phases. An analogous difference in $\Delta v$/$v(H)$ values for the two phases could also be observed in the case of SOMPT, but is hidden by the much stronger reverse peak due to magnetic fluctuations \cite{LUTHI1970,Moran,Petrova}.    

The results of ultrasonic measurements have brought new insights into magnetoelastic phenomena in magnetic materials, e.g. phenomena associated with phase transitions induced by a magnetic field in the uniaxial ferromagnet U$_{3}$Cu$_{4}$Ge$_{4}$ \cite{Gorbunov}, the ferrimagnet Dy$_{2}$Fe$_{14}$Si$_{3}$ \cite{Andreev}, the antiferromagnet UIrSi$_{3}$ \cite{Gaid2022}, or the itinerant metamagnets UTe$_{2}$ \cite{Valiska} and UCoAl \cite{Yoshizawa, Valiska}. In these cases, a single-step anomaly in $\Delta v$/$v(H)$ dependence at the critical point indicates FOMPT. Above a specific temperature, the inverted-peak anomaly associated with SOMPT emerges and remains up to $T_{C}$ in U$_{3}$Cu$_{4}$Ge$_{4}$  \cite{Gorbunov} and ferrimagnet Dy$_{2}$Fe$_{14}$Si$_{3}$ and up to $T_{N}$ in UIrSi$_{3}$. The evolution of the $\Delta v$/$v(H)$ anomaly with increasing temperature in itinerant metamagnets UTe$_{2}$  and UCoAl is connected with phenomena \cite{Yoshizawa, Valiska}, which are beyond the scope of this paper. 

U$_{3}$Cu$_{4}$Ge$_{4}$  is a uniaxial ferromagnet that displays a FM – PM transition in transverse fields [1] similar to our case of EuAl$_{12}$O$_{19}$. However, it has almost two orders of magnitude higher $T_{C}$ (= 73 K),  much stronger anisotropy ($\mu_{0}H_{c}$ at 2 K = 25 T) than EuAl$_{12}$O$_{19}$. Below 50 K, it exhibits FOMPT, accompanied by a simple stepwise anomaly with hysteresis. Above 50 K, the inverted-peak anomaly suddenly emerges, indicating the SOMPT, and remains up to $T_{C}$.

Given the above arguments, we conclude that the sharp dip at $H_{c}$ in $\Delta v$/$v(H)$ dependencies measured in EuAl$_{12}$O$_{19}$ at temperatures down to 0.32 K (see Figs. 2 and 4) manifests the second-order phase transition. Since the magnetocrystalline anisotropy of EuAl$_{12}$O$_{19}$ is very weak, we expect the appearance of eventual FOMPT at much lower temperatures.  

\section{Conclusions}
The magnetic phase transition induced by a magnetic field applied transverse to the spontaneous magnetization in the uniaxially anisotropic ferromagnet EuAl$_{12}$O$_{19}$ has been characterized by sound velocity measurements. The dramatic softening of the shear modulus $C_{44}$ at he critical field of the transition is manifested by the sharp dip in the field dependence of the ultrasound velocity. This result is interpreted as an effect of critical slowing of magnetic fluctuations at a second-order magnetic phase transition transferred to the lattice via magnetoelastic coupling. This feature was successfully modeled using magnetization data, confirming that the magnetoelastic coupling comes mainly from the exchange striction mechanism. This work establishes the possibility of modeling the magnetoelastic behavior of anisotropic ferromagnets under a transverse magnetic field based on the simple example of EuAl$_{12}$O$_{19}$, and it may be extended to other anisotropic ferromagnets. 

\section*{acknowledgments}
Crystal growth and some of the measurements of physical properties were carried out in the MGML (http://mgml.eu/), which is supported within the program of the Czech Research Infrastructures (project no. LM2023065). 

We acknowledge support from the Deutsche Forschungsgemeinschaft (DFG) through SFB 1143 (Project No.\ 247310070) as well as the support of the HLD at HZDR, member of the European Magnetic Field Laboratory (EMFL). We further acknowledge support under the European Union’s Horizon 2020 research and innovation program through the ISABEL project (No. 871106).

Project Quantum materials for applications in sustainable technologies (QM4ST), funded as project no. CZ.02.01.01/00/22\textunderscore008/0004572 by P JAK, call Excellent Research.

This work was supported by the Ministry of Education, Youth and Sports of the Czech Republic through the INTER-EXCELLENCE II program (LUABA24056).

\bibliography{apssamp}
\end{document}